\documentclass[twocolumn,showpacs,preprintnumbers,amsmath,amssymb]{revtex4}

\usepackage{graphicx}
\usepackage{dcolumn}
\usepackage{bm}

\begin{document}

\preprint{APS/123-QED}

\title{Magnetic properties of a spin-1/2 octagonal lattice}

\author{Satoshi Morota$^{1}$, Takanori Kida$^{2}$, Masayuki Hagiwara$^{2}$, Yasuyuki Shimura$^{3}$, Yoshiki Iwasaki$^{4}$, and Hironori Yamaguchi$^{1}$}

\affiliation{
$^1$Department of Physics, Osaka Metropolitan University, Osaka 599-8531, Japan\\
$^2$Center for Advanced High Magnetic Field Science (AHMF), Graduate School of Science, Osaka University, Osaka 560-0043, Japan\\
$^3$Graduate School of Advanced Science and Engineering, Hiroshima University, Hiroshima 739-8530, Japan\\
$^4$Department of Physics, College of Humanities and Sciences, Nihon University, Tokyo 156-8550, Japan
}


Second institution and/or address\\
This line break forced

\date{\today}

\begin{abstract}
We successfully synthesized a verdazyl-based complex, ($p$-Py-V-$p$-CN)$_2$[Cu(hfac)$_2$].
Molecular orbital calculations reveal that three types of antiferromagnetic (AF) interactions are involved in the formation of a spin-1/2 distorted octagonal lattice composed of the verdazyl radical and Cu spins.
The magnetic properties of the compound exhibited contributions from AF correlations and a phase transition to an AF ordered state at approximately $T_{\rm{N}}$ = 2.5 K.
Below $T_{\rm{N}}$, we observed a $T^2$ dependence of the specific heat, indicating dominant two-dimensional AF correlations within the octagonal lattice.
The magnetization curve in the low-temperature region exhibited a low-field linear increase, subsequent bending at 1/3 magnetization, and high-field nonlinear increase.
We reproduced the observed unique magnetic behavior through the numerical analysis based on the octagonal lattice.
These results demonstrate that the present compound exhibits magnetic properties characteristic of octagonal lattice topology.
\end{abstract}


\maketitle
\section{INTRODUCTION}
The type of plaquette, which is a minimal closed loop on the lattice, is a crucial feature that defines the topology of a spin system.
Recently, two-dimensional (2D) triangular and kagome lattices composed of triangular plaquettes with antiferromagnetic (AF) exchange interactions have attracted increasing interest owing to their degenerate ground states~\cite{sankaku1,sankaku2}. 
Two-dimensional square lattice antiferromagnets with a square plaquette have been extensively studied as the parent spin system of high-temperature superconductors, namely layered cuprates~\cite{square1}.
Furthermore, competing ferromagnetic and AF interactions within square lattices can induce frustration, demonstrating the stabilization of exotic quantum states~\cite{my_square1, my_square2, my_square3,my_square4}.
Pentagonal plaquettes with geometric frustration have also been subjects of interest~\cite{penta_1, penta_2, penta_3,penta_4}.
However, few model compounds with pentagonal plaquettes have been reported owing to the difficulty in stabilizing orbital coupling in inorganic materials. 
Experimental studies of organic radical compounds have recently demonstrated that low-symmetry structures in molecular-based systems are effective in forming spin lattices composed of pentagonal plaquettes~\cite{my_penta} 
Hexagonal plaquettes that form 2D honeycomb lattices have also gained increasing interest through recent studies on the Kitaev model, in which bond-dependent Ising-type interactions induce frustration~\cite{kita1}.
Because the honeycomb lattice has a small coordination number (i.e., 3), quantum fluctuations are significantly enhanced. 
Therefore, even slight perturbations, such as lattice distortion~\cite{dimer1, dimer2, dimer3} and randomness~\cite{uematsu,random}, can easily stabilize quantum states.
Spin lattices composed of polygonal plaquettes larger than hexagonal are also expected to exhibit enhanced quantum fluctuations due to their lower coordination numbers.
In addition, the incorporation of bond-dependent interactions similar to the Kitaev model, which can be realized by introducing transition metals with strong spin-orbit couplings, serves as a catalyst for further investigations in the realm of quantum physics.
However, research on such large polygonal systems remains unexplored both experimentally and theoretically.
In terms of octagonal plaquettes, which is the focus of this study, the topology of the octagonal lattice is expected to induce quantized magnetic behavior at 1/3 of the saturation magnetization.



Modulation of the molecular structures of organic radicals can generate advanced spin-lattice designs.
Utilizing the diversity of the molecular structure of triphenyl verdazyl radicals, we developed various verdazyl-based quantum organic materials (V-QOM) and demonstrated the realization of unconventional spin lattices that have not been realized in conventional inorganic materials~\cite{3Cl4FV, 3Br4FV, b26Cl2V, 3D_honeycomb, LM}.
Furthermore, we expanded V-QOM by combining them with transition metals.
By transforming verdazyl radicals into ligand structures, we synthesized radical-based complexes with 3$d$ transition metals. 
Moreover, by utilizing non-direct coordination to the radical center, we realized metal-radical exchange couplings comparable in energy scale to the intermolecular coupling between radicals, resulting in the formation of spin lattices comprising intramolecular $\pi-d$ couplings and intermolecular $\pi-\pi$ stacking~\cite{morotaMn, tsukiyama, morotaCo, 2DCo, Ni}.
In these complexes, the metal-radical $\pi-d$ couplings and the magnetic anisotropy and spin size inherent to the metallic elements are involved in the formation of spin lattices. 
Because the metal atoms are coordinated with two radical ligands in these complexes, each molecule contains interacted three spins, providing effective coupling units for forming spin lattices composed of hexagonal or larger plaquettes.

In this study, we successfully synthesized ($p$-Py-V-$p$-CN)$_2$[Cu(hfac)$_2$] ($p$-Py-V-$p$-CN = 3-(4-pyridinyl)-1-phenyl-5-
(4-cyanophenyl)-verdazyl, hfac = 1,1,1,5,5,5-hexafluoro-2,4-pentanedione), which is a verdazyl-Cu complex. 
Molecular orbital calculations revealed that three types of AF interactions are involved in the formation of a spin-1/2 distorted octagonal lattice composed of the verdazyl radical and Cu spins.
The magnetic properties exhibit contributions from AF correlations and a phase transition to an AF-ordered state at approximately $T_{\rm{N}}$ = 2.5 K.
Below $T_{\rm{N}}$, we observed a $T^2$ dependence of the specific heat, attributed to a linear dispersive mode in the 2D AF system.
The magnetization curve in the low-temperature region exhibited a low-field linear increase, subsequent bending at 1/3 magnetization, and high-field nonlinear increase.
From the numerical analysis using the quantum Monte Carlo (QMC) method, we explained the observed behavior characteristic of the topology of the octagonal lattice.

\section{EXPERIMENTAL}
We synthesized $p$-Py-V-$p$-CN via the conventional procedure for producing the verdazyl radical~\cite{verd}.
A solution of Cu(hfac)$_2$$\cdot$2H$_2$O (350.40 mg, 0.73 mmol) in 2 ml ethanol and 10 ml of heptane was refluxed at 60 $^\circ$C. 
A solution of $p$-Py-V-$p$-CN (473.75 mg, 1.4 mmol) in 5 ml of CH$_2$Cl$_2$ was slowly added and stirred for 1 h. 
After the mixed solution was cooled to room temperature, a dark-brown crystalline solid of ($p$-Py-V-$p$-CN)$_2$[Cu(hfac)$_2$] was separated by filtration and washed with heptane.
Single crystals were obtained via recrystallization from a mixed solvent of acetone and ethanol at 10 $^\circ$C.

The X-ray intensity data were collected using a Rigaku XtaLAB Synergy-S instrument.
The crystal structure was determined using a direct method using SIR2004 and refined using the SHELXL97 crystal structure refinement program.
Anisotropic and isotropic thermal parameters were employed for non-hydrogen and hydrogen atoms, respectively, during the structure refinement. 
The hydrogen atoms were positioned at their calculated ideal positions.
Magnetization measurements were conducted using a commercial SQUID magnetometer (MPMS-XL, Quantum Design).
The experimental results were corrected for the diamagnetic contribution, which are determined based on the numerical analysis to be described and confirmed to be close to that calculated by Pascal's method.
High-field magnetization in pulsed magnetic fields was measured using a non-destructive pulse magnet at AHMF, Osaka University.
The specific heat was measured using a commercial calorimeter (PPMS, Quantum Design) by using a thermal relaxation method down to approximately 0.45 K.
All the experiments utilized small, randomly oriented single crystals.

Molecular orbital (MO) calculations were performed using the UB3LYP method as broken-symmetry hybrid density functional theory calculations with basis sets 6-31G (intermolecule) and 6-31G($d$, $p$) (intramolecule). 
All calculations were performed using the GAUSSIAN09 software package.
The convergence criterion was set at 10$^{-8}$ hartrees.
We employed a conventional evaluation scheme to estimate the intermolecular exchange interactions in the molecular pairs~\cite{MOcal}. 

The QMC code is based on the directed loop algorithm in the stochastic series expansion representation~\cite{QMC2}. 
The calculations were performed for $N$ = 1200 under the periodic boundary condition, where $N$ denotes the system size.
The system showed no significant size-dependent effect.
All calculations were performed using the ALPS application~\cite{ALPS,ALPS3}.
To avoid the difficulty of calculations with site-dependent $g$ values, we used a uniform normalized $g$ value. 
Magnetic susceptibility results were calibrated using the average $g$ value.
Magnetization curve results were calibrated using the $g$ values corresponding to the predominant contributions from Cu spin and radical spin in the low- and high-field regions, respectively. 
The calibration was performed based on the field showing 1/3 magnetization saturation as the boundary.
The magnetization in the high-field region was adjusted to ensure a seamless connection with the data in the low-field region.

\section{RESULTS}
\subsection{Crystal structure and spin model}
Figure 1(a) shows the molecular structure of ($p$-Py-V-$p$-CN)$_2$[Cu(hfac)$_2$].
The crystallographic parameters are presented in Table I.
Both the verdazyl radical, $p$-Py-V-$p$-CN, and Cu$^{2+}$ ion possess a spin value of 1/2.
The Cu$^{2+}$ ion is coordinated with two $p$-Py-V-$p$-CN ligands, resulting in an octahedral coordination environment. 
The two radicals in the molecule are equivalent in terms of crystallography, owing to the presence of an inversion center at the Cu atom position.
The bond lengths and angles of the Cu atoms are listed in Table II.
Approximately 62 ${\%}$ of the total spin density is localized on the central ring consisting of four N atoms.
The phenyl and cyanophenyl rings account for approximately 13-17 ${\%}$ of the relatively large total spin density, whereas the pyridine ring accounts for less than 8 ${\%}$ of the total spin density.
MO calculations were performed to evaluate the exchange interactions.  
In the intermolecular case, two primary AF exchange interactions were identified between the radical spins. 
Their values are evaluated as $J_{\rm{V1}}/k_{\rm{B}}$ = 38 K and $J_{\rm{V2}}/k_{\rm{B}}$ = 29 K, defined within the Heisenberg spin Hamiltonian, given by $\mathcal {H} = J_{n}{\sum^{}_{<i,j>}}\textbf{{\textit S}}_{i}{\cdot}\textbf{{\textit S}}_{j}$, where $\sum_{<i,j>}$ denotes the sum over neighboring spin pairs.
The molecular pairs associated with both interactions are related by inversion symmetry, as shown in Figs. 1(b) and 1(c).
In the intramolecular case, an AF exchange interaction $J_{\rm{Cu}}/k_{\rm{B}}$ = $41$ K was evaluated between the spins on the radical and Cu atom.
Because MO calculations overestimate the intramolecular interactions between verdazyl radicals and transition metals~\cite{morotaMn, tsukiyama, Ni}, the actual value of $J_{\rm{Cu}}$ is smaller than that obtained from the MO evaluation.
Two MO couplings between radicals form one-dimensional structures, which are connected through the intramolecular $\pi$-$d$ coupling between radical and Cu spins, yielding a two-dimensional structure in the (110) plane, as shown in Fig. 1(d).
We assigned different spin sites, $S_{\rm{V}}$ and $S_{\rm{Cu}}$, based on the differences in the $g$ values.
From a topological perspective, the spin network is equivalent to a distorted octagonal lattice, as shown in Fig. 1(e).
Regarding interplane coupling, a weak but finite AF interaction, leading to the phase transition to the ordered state, was found between the radical spins and was estimated to be less than approximately 1/30 of $J_{\rm{V1}}$.


\begin{table}
\caption{Crystallographic data for ($p$-Py-V-$p$-CN)$_2$[Cu(hfac)$_2$].}
\label{t1}
\begin{center}
\begin{tabular}{cc}
\hline
\hline 
Formula & C$_{50}$H$_{32}$CuF$_{12}$N$_{12}$O$_{4}$\\
Crystal system & Triclinic \\
Space group & $P\bar{\rm{1}}$ \\
Temperature (K) & 100 \\
$a$ $(\rm{\AA})$ & 9.0483(4) \\
$b$ $(\rm{\AA})$ & 10.2948(5)  \\
$c$ $(\rm{\AA})$ & 14.0515(6)  \\
$\alpha$ (degrees) & 78.230(4) \\
$\beta$ (degrees) & 76.121(4)\\
$\gamma$ (degrees) & 74.005(4)\\
$V$ ($\rm{\AA}^3$) & 1208.09(9) \\
$Z$ & 1 \\
$D_{\rm{calc}}$ (g cm$^{-3}$) & 1.590\\
Total reflections & 2979 \\
Reflection used & 2661 \\
Parameters refined & 358 \\
$R$ [$I>2\sigma(I)$] & 0.0455  \\
$R_w$ [$I>2\sigma(I)$] & 0.1218 \\
Goodness of fit & 1.072 \\
CCDC & 2305915\\
\hline
\hline
\end{tabular}
\end{center}
\end{table}

\begin{table}
\caption{Bond distances ($\rm{\AA}$) and angles ($^{\circ}$) related to the Cu atom for ($p$-Py-V-$p$-CN)$_2$[Cu(hfac)$_2$].}
\label{t1}
\begin{center}
\begin{tabular}{cc@{\hspace{1.5cm}}cc}
\hline
\hline
Cu--N1 & 2.00 & O1--Cu--O2 & 86.2 \\
Cu--N2 & 2.00 & O2--Cu--O3 & 93.8 \\
Cu--O1 & 1.98 & O3--Cu--O4 & 86.2\\
Cu--O2 & 2.32 & O4--Cu--O1 & 93.8 \\
Cu--O3 & 1.98 & N1--Cu--O4 & 93.9 \\
Cu--O4 & 2.32 & O4--Cu--N2 & 86.1\\
 &  & N2--Cu--O2 & 93.9 \\
 &  & O2--Cu--N1 & 86.1 \\
 &  & N1--Cu--O3 & 89.5 \\
 &  & O3--Cu--N2 & 90.6 \\
 &  & N2--Cu--O1 & 89.5 \\
 &  & O1--Cu--N1 & 90.6 \\
\hline
\hline
\end{tabular}
\end{center}
\end{table}

\begin{figure*}[t]
\begin{center}
\includegraphics[width=38pc]{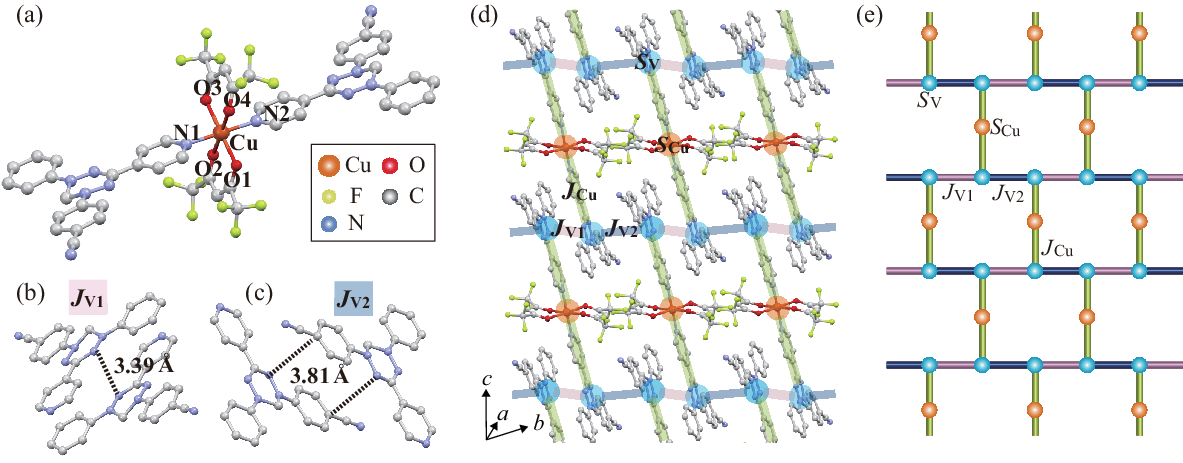}
\caption{(color online) (a) Molecular structure of ($p$-Py-V-$p$-CN)$_2$[Cu(hfac)$_2$], which causes intramolecular exchange interaction $J_{\rm{Cu}}$ between radical and Cu spins. 
The hydrogen atoms have been omitted for clarity. 
Molecular pairs associated with the exchange interactions of (b) $J_{\rm{V1}}$  and (c) $J_{\rm{V2}}$, respectively.
The dashed lines indicate N-N and N-C short contacts. 
(d) Crystal structure forming a 2D spin lattice in the (110) plane.
The blue and orange nodes represent the spin-1/2 on the radical and Cu ion, $S_{\rm{V}}$ and $S_{\rm{Cu}}$, respectively.
The thick lines represent the exchange interactions $J_{\rm{V1}}$, $J_{\rm{V2}}$, and $J_{\rm{Cu}}$.
(e) Corresponding spin-1/2 distorted octagonal lattice.
}
\end{center}
\end{figure*}

\begin{figure}[t]
\begin{center}
\includegraphics[width=20pc]{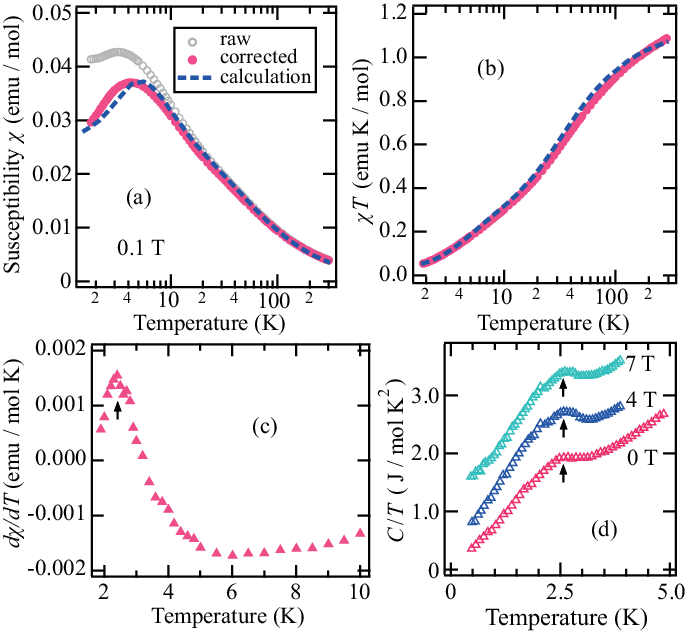}
\caption{(color online) Temperature dependence of (a) magnetic susceptibility ($\chi$ = $M/H$), (b) $\chi T$, and (c) $d\chi/dT$ of ($p$-Py-V-$p$-CN)$_2$[Cu(hfac)$_2$] at 0.1 T. 
The open circles denote raw data, and the closed circles are corrected for the paramagnetic term due to the impurity.
The broken lines represent the calculated results for the spin-1/2 distorted octagonal lattice with $\alpha=J_{\rm{V2}}/J_{\rm{V1}}$ = 0.86 and $\beta =J_{\rm{Cu}}/J_{\rm{V1}}$ = 0.39 ($J_{\rm{V1}}/k_{\rm{B}}$ = 43 K).
(d) Temperature dependence of the specific heat $C/T$ of ($p$-Py-V-$p$-CN)$_2$[Cu(hfac)$_2$] at 0, 4, and 7 T. 
For clarity, the values for 4 and 7 T have been shifted up by 0.6 and 1.4 J/ mol K$^2$, respectively.
The arrows indicate the phase transition temperature $T_{\rm{N}}$.
}\label{f1}
\end{center}
\end{figure}

\begin{figure}[t]
\begin{center}
\includegraphics[width=19pc]{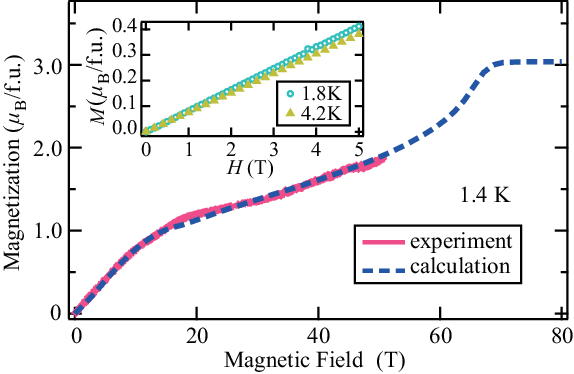}
\caption{(color online) Magnetization curve of ($p$-Py-V-$p$-CN)$_2$[Cu(hfac)$_2$] at 1.4 K in pulsed magnetic fields.
The broken line represents the calculated result for the spin-1/2 distorted octagonal lattice with $\alpha=J_{\rm{V2}}/J_{\rm{V1}}$ = 0.86 and $\beta =J_{\rm{Cu}}/J_{\rm{V1}}$ = 0.39 ($J_{\rm{V1}}/k_{\rm{B}}$ = 43 K). 
The inset shows magnetization curves at 1.8 K and 4.2 K in static magnetic fields.
}\label{f3}
\end{center}
\end{figure}

\begin{figure}[t]
\begin{center}
\includegraphics[width=20pc]{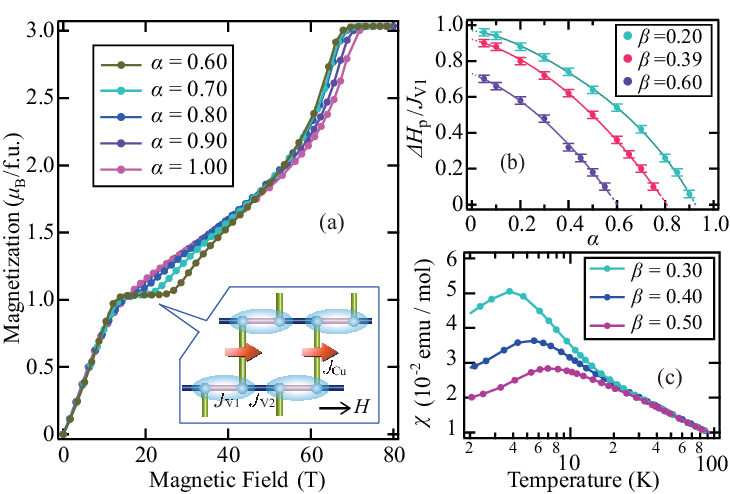}
\caption{(color online) 
(a) Calculated magnetization curves at $T/J_{\rm{V1}}$ = 0.035 for the spin-1/2 distorted octagonal lattice with the representative values of $\alpha=J_{\rm{V2}}/J_{\rm{V1}}$ with fixed $\beta =J_{\rm{Cu}}/J_{\rm{V1}}$ = 0.39. 
The illustration shows a schematic picture of the quantum state in the 1/3 plateau phase, where the ovals and arrows represent the valence bond singlet of $S_{\rm{V}}$ via $J_{\rm{V1}}$ and $S_{\rm{Cu}}$ polarized in the external field direction, respectively.
(b) $\alpha$ dependence of normalized field range of the 1/3 magnetization plateau for $\beta$ = 0.20, 0.39 (this compound), and 0.60.
The lines are guides for the eye.
(c) Calculated magnetic susceptibilities for the spin-1/2 distorted octagonal lattice with the representative values of $\beta$ with fixed $\alpha$ = 0.86. 
}\label{f3}
\end{center}
\end{figure}

\subsection{Magnetic and thermodynamic properties}
Figure 2(a) shows the temperature dependence of the magnetic susceptibility $\chi$ at 0.1 T.
Assuming the conventional paramagnetic behavior $C_{\rm{imp}}/T$, where $C_{\rm{imp}}$ is the Curie constant of the spin-1/2 impurities, we evaluated the paramagnetic impurities to be approximately 1.8 ${\%}$ of all spins, which is defined to fit the following calculated result. 
We observe a broad peak in the low-temperature region, indicating dominant AF contributions. 
Correspondingly, the temperature dependence of $\chi T$ indicates a decrease with decreasing temperatures, as shown in Fig. 2(b).
The decrease in $\chi T$ becomes gradual below approximately 20 K.
If a significant difference in the magnitude of exchange interactions exists, an energy separation occurs owing to the difference in the temperature region where the correlation becomes dominant, yielding a multistep change in $\chi T$~\cite{my_penta, b26Cl2V, tsukiyama}.
Accordingly, the observed change in the decreasing trend of $\chi T$ is expected to reflect the differences in the exchange interactions, i.e., the distortion in the octagonal lattice.
Furthermore, we observed a distinct sharp peak in the temperature derivative of $\chi$ ($d \chi$/$dT$) at approximately 2.5 K, as shown in Fig. 2(c), which is a signal indicating a phase transition to an ordered state induced by weak but finite interplane couplings.

Figure 2(d) shows the temperature dependence of the specific heat $C/T$.
At zero field, an anomaly at approximately $T_{\rm{N}}$ = 2.5 K can be observed, which is consistent with the peak temperature of $d \chi$/$dT$ and indicates the phase transition to the ordered state. 
The peak signal remains almost unchanged up to 7 T, suggesting the existence of strong AF exchange couplings of several tens of Kelvins, which is consistent with the MO evaluations. 
Below $T_{\rm{N}}$, $C/T$ shows a $T$-linear decrease, which corresponds to a $T^2$ dependence of the specific heat attributed to a linear dispersive mode in the 2D AF system.
Despite the weak interplane couplings that can cause a phase transition to the ordered state, the present spin system has a 2D character, yielding dominant 2D correlations within the octagonal lattice.

Figure 3 presents the magnetization curve at 1.4 K under a pulsed magnetic field.
A significant bending is observed at approximately 18 T, where the magnetization corresponds to 1/3 of the saturation value.
In the low-field region below the bending, the magnetization exhibits an almost linear increase, as shown in the inset of Fig. 3. 
Because the phase transition occurs at $T_{\rm{N}}$ = 2.5 K, this behavior does not significantly depend on the presence or absence of the magnetic order, indicating that the linear increase is an intrinsic behavior originating from the 2D correlations in the octagonal lattice.   
Meanwhile, we observe a nonlinear increase in the high-field region above the bending, which is typical in quantum spin systems.


\section{Analyses and Discussion}
Considering the results of MO calculations, the magnetic properties were investigated based on the spin-1/2 octagonal lattice composed of the AF interactions, $J_{\rm{V1}}$, $J_{\rm{V2}}$, and $J_{\rm{Cu}}$. 
Using the QMC method, we calculated the magnetic susceptibility and magnetization curves by considering the parameters $\alpha=J_{\rm{V2}}/J_{\rm{V1}}$ and $\beta =J_{\rm{Cu}}/J_{\rm{V1}}$.
First, we note the bending of the magnetization curve at approximately 1/3 of the magnetization saturation value. 
Regarding the coordination number of each spin site, the lower coordination number of $S_{\rm{Cu}}$ is expected to enhance the effect of polarization by the external magnetic fields. 
Consequently, below the bending field, the polarization of the magnetic moment is attributed to $S_{\rm{Cu}}$, reaching an almost fully polarized state at the bending field.
The full polarization of $S_{\rm{Cu}}$ lacks sufficient degrees of freedom to alter the ground state.
Therefore, the magnetic behavior above the bending field mainly reflects the magnetic properties of the alternating chain formed by $J_{\rm{V1}}$ and $J_{\rm{V2}}$.
In certain parameter ranges, this effective alternating chain can create a singlet state with an energy gap, leading to the appearance of the 1/3 magnetization plateau, as shown in the inset of Fig. 4(a). 
The $\alpha$ dependence of the magnetization curves exhibits a 1/3 magnetization plateau attributed to the spin gap of the effective alternating chain, which becomes prominent as the alternating ratio increases, i.e., a smaller value of $\alpha$, as shown in Fig. 4(a).
In Fig. 4(a), the magnetic field for each $\alpha$ was scaled to have the same energy scale. 
Figure 4(b) shows the field range of the 1/3 magnetization plateau, ${\Delta}H_{\rm{p}}$, evaluated from the calculated magnetization curves.
For a fixed value of $\beta$, the plateau region decreases as $\alpha$ increases and eventually disappears even with some alternation ($\alpha$ ${\textless}$ 1). 
The plateau region also decreases as $\beta$ increases.
These parameter dependencies indicate that the plateau state becomes unstable as the two-dimensionality of the system increases.
In the low-field and low-temperature regions, the magnetic properties are mainly attributed to $S_{\rm{Cu}}$.
Accordingly, the broad peak of the magnetic susceptibility observed in the low-temperature region primarily originates from the AF correlation associated with $J_{\rm{Cu}}$, leading to a pronounced dependence on $\beta$.
Figure 4(c) indicates the calculated $\beta$ dependence of the broad peak, where each magnetic susceptibility is scaled to reproduce the experimental result in a higher temperature region. 


Based on these parameter dependencies, we obtained good agreement between the experimental and calculated results using $\alpha$ = 0.86 and $\beta$ = 0.39 ($J_{\rm{V1}}/k_{\rm{B}}$ = 43 K), as shown in Figs. 2(a), 2(b), and 3.
Assuming an isotropic $g$ value of 2.0 for the organic radicals, we determined an average $g$ value of approximately 2.07 for the Cu spins.
The calculated results successfully reproduced the main features of the magnetization curve, i.e., the low-filed linear increase, subsequent bending, and high-field nonlinear increase.
In the present octagonal lattice, the alternation of the $J_{\rm{V1}}$-$J_{\rm{V2}}$ chain is not large enough to stabilize the 1/3 plateau phase, as shown in Fig. 4(b), yielding only the bending at 1/3 magnetization. 
In the low-field region below the bending field, the magnetization largely reflects the polarization of $S_{\rm{Cu}}$ connected by the weakest $J_{\rm{Cu}}$.
Strong quantum fluctuations attributed to the AF correlations in the effective chain are also expected to emphasize the weak coupling between Cu spins.
Therefore, the low-field linear increase originates from the classical paramagnetic-like behavior of $S_{\rm{Cu}}$, resulting in the suppression of the nonlinear increase characteristic of quantum spin systems. 
In the higher-filed region, the correlations of the effective $J_{\rm{V1}}$-$J_{\rm{V2}}$ chain become dominant, yielding strong quantum fluctuations owing to its one-dimensionality.
Consequently, the magnetization curve exhibits a nonlinear increase towards saturation.
Considering that the symmetry of the octagonal lattice allows the formation of a magnetic state based on 1/3 magnetization, and the enhanced quantum fluctuations are primarily attributed to the small coordination number of the octagonal lattice, the observed magnetization curve demonstrates the quantum behavior specific to the octagonal topology.


\section{Summary}
In this study, we synthesized a verdazyl-based complex, ($p$-Py-V-$p$-CN)$_2$[Cu(hfac)$_2$].
The MO calculations revealed three AF interactions forming a spin-1/2 distorted octagonal lattice composed of the verdazyl radical and Cu spins.
The temperature dependence of the magnetic susceptibility exhibited a broad peak, showing dominant AF contributions. 
The temperature dependence of the specific heat exhibited an anomaly at approximately $T_{\rm{N}}$ = 2.5 K, showing a phase transition to an AF-ordered state.
Below $T_{\rm{N}}$, we observed a $T^2$ dependence of the specific heat attributed to a linear dispersive mode in the 2D AF system, indicating that the present spin system has dominant 2D correlations within the octagonal lattice.
The magnetization curve exhibited an almost linear increase and subsequent bending at 1/3 magnetization.
A nonlinear increase was observed in the high-field region above the bending, which is a typical behavior for spin systems with strong quantum fluctuations.
The magnetic properties of the verdazyl-based complex were investigated based on the spin-1/2 octagonal lattice using the QMC method, and good agreement between the experimental and calculated results were obtained.
In the octagonal lattice, the lattice distortion is not large enough to stabilize the 1/3 plateau phase, yielding only the bending at 1/3 magnetization. 
In the low-field region below the bending, the linear increase in magnetization originates from the classical paramagnetic-like behavior of weakly coupled spins.
On the other hand, in the high-field region above the bending, enhanced quantum fluctuations attributed to the effective 1D correlations induce the nonlinear increase in magnetization.

In this study, we demonstrated that the present compound exhibits magnetic properties characteristic of the topology of an octagonal lattice.
Moreover, using V-QOM is highly effective for designing spin lattices composed of large polygonal plaquettes, whose unique topologies may give rise to unconventional quantum phenomena.

\begin{acknowledgments}
We thank Y. Hosokoshi for letting us use the laboratory equipments.
This research was partly supported by KAKENHI (Grant No. 23K13065) and JST FOREST (No. JPMJFR2233).
A part of this work was performed under the interuniversity cooperative research program of the joint-research program of ISSP, the University of Tokyo.
\end{acknowledgments}


\end{document}